\documentclass[12pt]{article}
\usepackage{amsfonts}
\usepackage{amsmath, graphicx, epsfig,psfrag, verbatim}
\usepackage{eucal,  amssymb,amsmath,amsthm,graphicx, amsfonts, latexsym}
\usepackage[utf8]{inputenc}

\newcommand{\hfigwidth}{0.45\textwidth}

\begin{document} \parskip=5pt plus1pt minus1pt \parindent=0pt

\title{Summer vacation and COVID-19: effects of metropolitan people going to summer provinces}

\author{Tom Britton$^{1}$ and Frank Ball$^{2}$}
\date{\today}
\maketitle

\begin{abstract}
Many countries are now investigating what the effects of summer vacation might be on the COVID-19 pandemic. Here one particular such question is addressed: what will happen if large numbers of metropolitan people visit a less populated province during the summer vacation? By means of a simple epidemic model, allowing for both short and long-term visitors to the province, it is studied which features are most influential in determining if such summer movements will result in large number of infections among the province population. The method is applied to the island of Gotland off the South East coast of Sweden. It is shown that the amount of mixing between the metropolitan and province groups and the fraction of metropolitan people being infectious upon arrival are most influential. Consequently, minimizing events gathering both the province and metropolitan groups and/or reducing the number of short-term visitors could substantially decrease spreading, as could measures to lower the fraction initially infectious upon arrival.
\end{abstract}

\footnotetext[1]{Stockholm University, Department of Mathematics, Sweden. E-mail: tom.britton@math.su.se}
\footnotetext[2]{University of Nottingham, School of Mathematical Sciences, UK.}

\section*{Introduction}\label{sec-Intro}

Most countries in Europe and North America are now considering slowly relaxing restrictions aimed at reducing the spread of the SARS-COV2 virus. There are also substantial economical stakes for the tourist industry to have summer visitors even at some reduced level. A relevant question therefore is to study the effect of increased movements of people during the summer vacation period, typically going from metropolitan areas with higher mixing rates to provincial districts with fewer residents and lower mixing rates.

In this study we investigate such effects, focusing on a situation where a large number of individuals from one metropolitan area go to one provincial area during the summer vacation. It is assumed that the metropolitan people have higher mixing rates (hence also a higher reproduction number) when compared to the provincial community, but also that there is higher immunity level among the metropolitan people. The metropolitan group is separated into two categories, short-term visitors that typically stay in hotels/hostels and eat out more, and  long-term visitors that have their own summer homes (or long-term camping vans) and who mix less with the provincial community.

The aim of the paper is to investigate which factors are most influential in determining if there will be a substantial increase in disease spreading in the provincial community, or if only few additional infections will occur. After defining a general framework, we illustrate our methods by investigating the situation on the island of Gotland off the Swedish South East coast, which every year receives large numbers of both long and short-term visitors.

\section*{An epidemic model for metropolitan people moving to a province region during the summer}

We now describe briefly an epidemic model for the situation where a large number of metropolitan people move to a provincial region during the summer holidays. We start by determining the new reproduction number in the province after the metropolitan people have arrived. Then comes a more detailed model describing the dynamics of the epidemic over time.

\subsection*{Basic reproduction numbers}

We consider two separate populations, denoted $M$  (metropolitan) and $P$ (provincial). For simplicity we assume each of them to be homogeneously mixing and we consider a simple SIR (susceptible $\to$ infective $\to$ recovered) epidemic model within each community \cite{DHB13}, though as indicated in the SI these results also hold if an exposed (latent) period is introduced. The $M$-population has reproduction number $R_0^{(M)}$, restrictions/preventive measures with overall effect $c^{(M)}$ and at the start of summer vacation, a fraction $r^{(M)}$  are recovered and immune, a tiny fraction $\epsilon^{(M)}$ are infectious, and the remaining fraction, $1-r^{(M)}-\epsilon^{(M)}$, are susceptible. The effective reproduction number among metropolitan people, taking restrictions and immunity into account, hence equals $R_E^{(M)}=R_0^{(M)}(1-c^{(M)})(1-r^{(M)})$.

The corresponding quantities for the $P$-population are $R_0^{(P)}$, $c^{(P)}$ and $r^{(P)}$, with effective reproduction number $R_E^{(P)}=R_0^{(P)}(1-c^{(P)})(1-r^{(P)})$. We assume that $R_0^{(M)} > R_0^{(P)}$, and also that $r^{(M)}>r^{(P)}$. The typical case treated is where $c^{(M)}=c^{(P)}$, but this is not necessary.

During the summer period (assumed to have fixed start and end dates) a large number of metropolitan people visit the province. The visitors are of two types: long and short-term visitors -- a fraction $\alpha_L$ being long-term visitors and the remaining fraction $\alpha_S=1-\alpha_L$ being short-term visitors.
At the start of the summer vacation, a large group from the $M$-population goes to the province. Long-term visitors stay the whole summer season and short-term visitors stay shorter but upon leaving are replaced by new short-term visitors, thus keeping the number of short-term visitors constant. The relative size of the $M$-population that moves to the province compared to the whole summer population in the province is $\pi_M$ (so if this group is twice as large as the $P$-population then $\pi_M=2/3$). It is assumed that the mixing within the $M$-population that move to the province is unchanged, meaning that the internal effective reproduction number remains at $R_E^{(M)}$,  and similarly for the $P$-population.

Once the $M$-population arrives in the province there is an added contact rate between the $M$ and the $P$-populations (the within contact rates are unchanged). We denote the average number of close contacts an infected long-term $M$-individual has with $P$-individuals during his/her infectious period by $R_{M\to P}$. The corresponding average for short-term visitors equals $(1+k_S)R_{M\to P}$, so $k_S$ is the added mixing rate for short-term visitors compared with long-term visitors (the rational for short-term visitors having more contacts with $P$-people is that they typically stay in hotels, eat at restaurants et cetera). Close contacts are symmetric, so by taking population proportions into account this induces average number of contacts with long and short-term visitors that a $P$-individual has during the infectious period. We point out that these ``mixing'' reproduction numbers take preventive measures into account, but not immunity, so are best compared with $R_0^{(M)}c^{(M)}$ and $R_0^{(P)}c^{(P)}$, respectively. Let short-term M-individuals be type 1, long-term $M$-individuals type 2 and $P$-individuals type 3. Then it is shown in the SI that the average number of $j$-individuals a typical $i$-individual infects during their infectious period, denoted $m_{ij}$, for $i,j=1,2,3$ is given in the following next generation matrix:
\[
A=
\begin{pmatrix}
 R_E^{(M)}\alpha_S &  R_E^{(M)}\alpha_L&  (1+k_S) R_{M \to P}(1-r^{(P)})\\
 R_E^{(M)}\alpha_S &  R_E^{(M)} \alpha_L&  R_{M \to P}(1-r^{(P)})\\
 (1+k_S) R_{M \to P}(1-r^{(M)})\frac{\alpha_S \pi_M}{1-\pi_M}&  R_{M \to P}(1-r^{(M)})\frac{\alpha_L\pi_M}{1-\pi_M} &  R_E^{(P)}
\end{pmatrix}.
\]

The overall effective reproduction number $R_E$ for spreading in the provincial region at the start of the summer is given by the largest eigenvalue of this matrix. We refrain from giving the long expression for $R_E$ but it is easily computed numerically for any particular example.
If $R_E\le 1$ there will not be many infections during the summer, whereas if $R_E>1$ there is a risk for an outbreak, and the approximate size of such an outbreak can be determined numerically as described in the SI.  It can be shown (see SI) that if $R_E^{(M)}<1$ and $R_E^{(P)}<1$, then $R_E\le 1$ if and only if
\[
R_{M \to P} \le \sqrt{\frac{(1-R_E^{(M)})(1-R_E^{(P)})(1-\pi_M)}
{(1-r^{(P)})(1-r^{(M)})\pi_M[(1+k_S \alpha_S)^2+\alpha_S(1-\alpha_S)(1-R_E^{(M)})]}}.
\]
As a consequence, the closer $R_E^{(M)}$ and $R_E^{(P)}$ are to 1, the less mixing between $M$ and $P$-individuals can be allowed in order to keep $R_E\le 1$.

\subsection*{Time dynamic considerations with short and long term visitors}

We assume that summer season lasts for $s_V$ days. During this period the long-term visitors remain in the province whereas the short-term visitors leave after on average $1/\mu_S$ days, implying that $\mu_S$ is the rate at which short-term visitors leave. The number of short-term visitors remains unchanged, and their fraction compared to the whole summer population in the province equals $\alpha_S \pi_M$. It is assumed that the contact rates, the preventive measures and the immunity levels for the short and long-term visitors arriving to the province remain constant during the summer.

We choose a deterministic SEIR (susceptible $\to$ exposed $\to$ infective $\to$ recovered) epidemic model \cite{DHB13} with on average a $1/\nu$ day exposed (latent) period and a $1/\gamma$ day infectious period. (A deterministic model approximates well a more realistic stochastic model as shown in the SI). The rate at which a typical $M$-individual mixes with other $M$-individuals (taking preventive measures into account) equals $\lambda^{(M)}= R_0^{(M)} (1-c^{(M)})\gamma$, and similarly for the $P$-individuals:
$\lambda^{(P)}= R_0^{(P)} (1-c^{(P)})\gamma$. Only contacts with susceptible individuals result in infections -- other contacts have no effect. The rate at which a $P$-individual has contact with $M$-individuals differs between short and long-term visitors. Long-term visitors have contact with $M$-individuals at rate $R_{M\to P}\gamma$ and short-term visitors have contact with $M$-individuals at rate $(1+k_S) R_{M\to P}\gamma$. Viewed from the perspective of a $P$-individual, he/she will have contact with short-term $M$-individuals at rate $R_{M\to P}\gamma\pi_M \alpha_S(1+k_S)/(1-\pi_M)$ and with long-term $M$-individuals at rate $R_{M\to P}\gamma\pi_M (1-\alpha_S)/(1-\pi_M)$. Details of the model and its defining set of differential equations are given in the SI.

The model gives what fractions are susceptible, exposed, infectious and immune, respectively, for short and long-term $M$-individuals as well as $P$-individuals over time. The main focus is to study how many people in the province become infected as a result of the $M$-individuals visiting the province.

\section*{Illustration: Stockholmer's visiting the island of Gotland}

We illustrate our result by considering the island of Gotland off the South East coast of Sweden, which every year receives many visitors, in particular from the Stockholm region. To simplify the illustration we assume all visitors are from the Stockholm region. The total population of Gotland is 60~000 individuals. During the summer, from late June to early August, around 500~000 visitors come to Gotland. Most of them visit for  up to 1 week (short-term visitors), but around 100~000 people reside in summer houses or long-term camping homes in Gotland, typically staying a month (long-term visitors). We assume the long-term $M$-visitors remain on the island for the entire summer season, i.e.~for the entire $s_V =35$ days (5 weeks), whereas the short-term $M$-visitors remain on average for $1/\mu_S=4$ days. On a given day during the summer season (late June to early August) approximately 150~000 visitors are on Gotland (so about 50~000 short-term visitors).  Further, we assume that the short-term $M$-visitors have double the interaction with $P$-individuals due to more contacts at restaurants, hotels and similar, so $k_S=1$.

In our base model it is assumed that the basic reproduction number, effects of preventive measures and immunity level among the Stockholmer’s are $R_0^{(M)}=2.2$, $c^{(M)}=0.5$ and $r^{(M)}=0.3$, and hence effective reproduction number $R_E^{(M)}=2.2*0.5*0.7=0.77$ in late June (the estimate in late May was 0.85 but is expected to continue declining \cite{FHM20}).
The corresponding numbers for the people of Gotland are set to $R_0^{(P)}= 1.1$, $c^{(P)}=0.5$ and $r^{(P)}=0.05$, implying that $R_E^{(P)}=0.522$ (the basic reproduction number was chosen as half of that in Stockholm, same preventive measures, and a small fraction being immune in late June). The fraction of the summer population coming from Stockholm equals $\pi_M=15/21$, with a fraction $\alpha_S=2/3$ of these being short-term visitors. Finally, $R_{M\to P}$ was set to $0.375$ to be compared with $R_0^{(M)}(1-c^{(M)})=1.1$, so the amount of close contacts long-term metropolitans have with provincial people is about 1/3 compared to contacts with other metropolitans (including family). For these quantities we study how many Gotland people would get infected depending on the amount of interaction between the $M$ and the $P$ individuals. We also vary other input parameters to see which parameters have most influence on the results.

We assume further that the exposed period lasts on average $1/\nu=5$  days and is followed by an infectious period of on average $1/\gamma =4$ days (in approximate agreement with \cite{FMG20}). The summer season starts on June 25 with 0.5\% of the $M$-individuals and 0.1\% of the $P$-individuals being infectious, the former is also varied in a sensitivity analysis.

\section*{Results}

\begin{figure}[ht]
\begin{center}
\includegraphics[scale=0.9, angle=0]{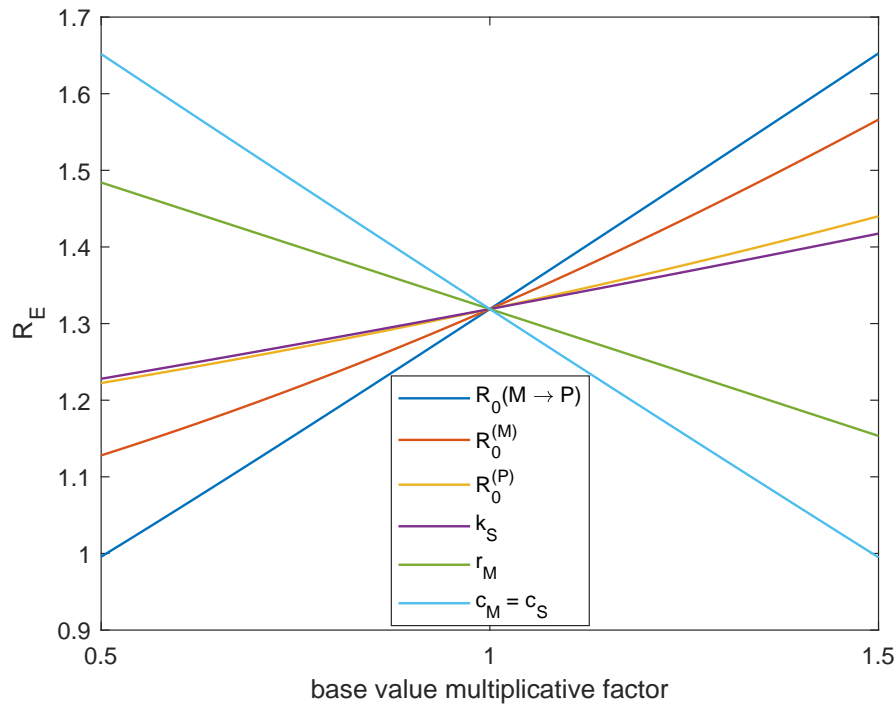}
\end{center}
\caption{Plot of $R_E$ where each of 6 parameters are varied up and down 50\%. }
\label{Fig_R0}
\end{figure}

We first compute the effective reproduction  number $R_E$ for the province in the beginning of the summer, the largest eigenvalue of the matrix $A$ defined above. For the base values described above we have $R_E=1.32$. In Figure \ref{Fig_R0} we plot $R_E$ for the base values, but where we vary each of the six most influential model parameters between 50\% of the base value up to 150\% of the base value, thus creating six curves.

From the figure it is seen clearly that the mixing between the $M$ and $P$-individuals is very influential: the more they mix the higher potential for transmission. Another important factor is the effect $c$ of the set of preventive measures, and, naturally, the higher effect of preventive measures the \emph{lower} is $R_E$. The level of immunity among $M$-individuals, $r_M$, and the reproduction number among $P$-individuals also play a significant role. The parameters having smallest effect on $R_E$ are the reproduction number within $P$-individuals and the added mixing rate for short-term visitors. It appears like $R_E$ depends linearly on each parameter but this is not the case as would be apparent if the range of variation is increased.

Next we study how many $P$-individuals  might get infected as a consequence of the $M$-individuals visiting the province during the summer, also varying the more influential model parameters. More precisely, we consider the
fraction of $P$-individuals that have been infected 100 days after the summer season begins, assuming that all metropolitan visitors leave after 35 days.

\begin{table}[ht]
\caption{The additional fraction infected provincial people after summer season, varying each parameter up and down 50\%. Fraction infected with all parameters at base values gives 5.1\% infected}
\centering 
\begin{tabular}{| c | c | c | c | c | c|c | c |}
\hline
Relative effect & $R_0^{(M)}$ & $R_0^{(P)}$ & $R_{M\to P}$ & $k_S$ & $r_M$ & $c_M=c_S$ & $\epsilon_M$
\\
\hline                  
Down (so 50\%) & 4.0 & 3.3 & 2.3 & 4.3 & 6.0 & 12.2 & 2.8
\\
Up (so 150\%) & 6.8  & 9.4 & 9.4 & 6.0 & 4.3 & 2.6 & 7.2
\\
[1ex]      
\hline 
\end{tabular}\label{tab_frac_inf}
\end{table}
\bigskip
The actual numerical values are not our main focus, but the effect of altering different parameters carries important information. In Table \ref{tab_frac_inf} it is seen that $R_{M\to P}$ and $R_0^{(P)}$, followed by the fraction of $M$-individuals being infectious upon arrival, $\epsilon_M$, have the greatest impact on the final fraction getting infected.  It may seem surprising that $R_0^{(P)}$ is more influential than $R_0^{(M)}$, which is the opposite of that the effective reproduction number $R_E$ in Figure~\ref{Fig_R0}, however, as explained in the SI, this is primarily owing to further spread in the $P-$population after all metropolitan visitors have
departed, when the value of $R_0^{(M)}$ is clearly immaterial.  A secondary factor, also explained in the SI, is
that $R_E$ is an indicator of overall spread, while Table \ref{tab_frac_inf} focuses on just the provincial population.

\begin{figure}[ht]
\begin{center}
\includegraphics[scale=0.9, angle=0]{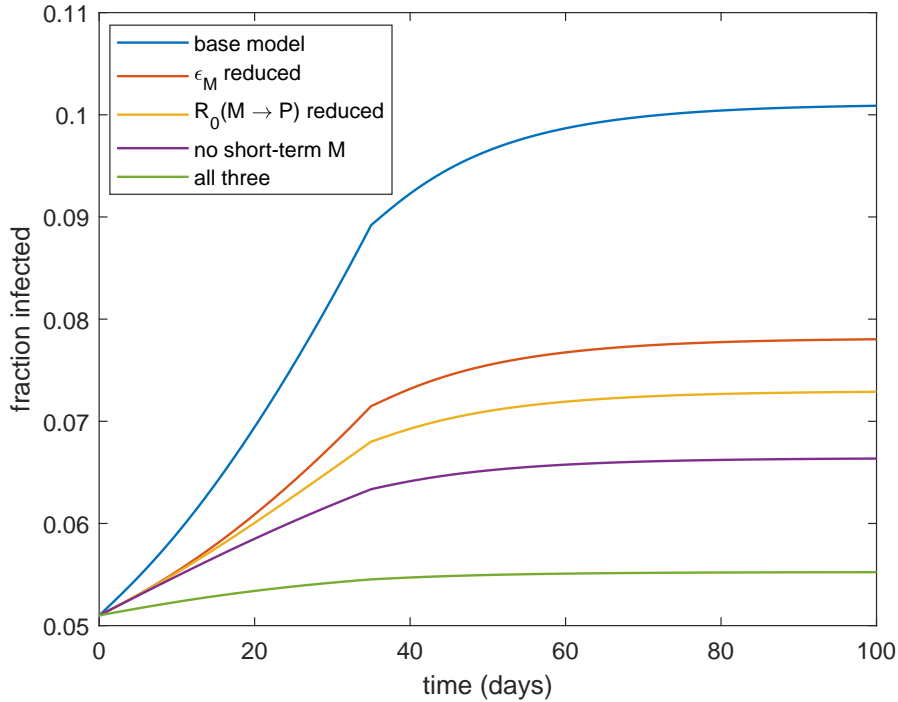}
\end{center}
\caption{Plot of cumulative fraction infected among provincial people over time (starting with 5\% immunity) for base model (blue), 50\% reduction in fraction of metropolitans being infectious on arrival (red),  50\% reduced contacts between metropolitan and provincial people (yellow), stopping short-term visitors (purple), and all three. Metropolitans visit province for 35 days, after this only province people remain.  }
\label{Fig_timeevolve}
\end{figure}

Finally, in Figure \ref{Fig_timeevolve} we plot the time evolution of the epidemic starting with the arrival of the $M$-people arriving to the province.
The curve plots the fraction of $P$-individuals that have been infected over time (assuming 5\% at the start of the summer season, day 0). Short-term visitors remain on average 4 days and are replaced by new short-term visitors when leaving, and all $M$-individuals leave the province after the summer season is over on day 35. The blue curve is for the base model, and the other curves show the same quantity, but where the mixing between $M$ and $P$-individuals is reduced by 50\% (yellow), where short-term visitors are not allowed (purple), and where the fraction of $M$-individuals being infectious upon arrival is reduced by 50\% (red), perhaps by postponing the season opening. The lowest curve shows the combined effect of all three measures.

\section*{Conclusions and discussion}

The paper treats a simple epidemic model, with the explicit aim of evaluating the effect of a large number of metropolitan people, with higher reproduction number and immunity, going to a provincial district having lower reproduction number and immunity level. The model is over-simplified and some of its parameters are hard to give correct numerical values (in particular mixing rates between metropolitan and provincial individuals), so the exact numerical values describing e.g.\ how many that will get infected among the provincial group, should not be relied upon. However, we do believe that some important qualitative findings come out of the analysis, in particular which features/parameters have highest impact on the studied outcomes. It is assumed implicitly that transmission is low in the province and declining in the metropolitan area. If this was not the case, having large number of metropolitan people visiting the province may have severe consequences.

In the illustration for the island of Gotland it is seen that the most influential parameter on how many that will get infected, is the amount of mixing between provincial people and the visitors. Reducing such mixing, for example by reducing the number of visitors having many contacts with people from the province (typically short-term visitors), or stopping events where there is substantial mixing between metropolitan and province people will hence reduce spreading the most. Another key determinant is how many of the visitors carry the virus upon arrival. If transmission has gone down significantly in the metropolitan region before the summer season starts, this will really help in keeping the number province people getting infected at a low level. A third important observation is that relaxing preventive measures at the same time as visitors come to the province could have clear negative effects.

The model contains several assumptions which could have been dealt with differently. For example, the model has no age structure, no heterogeneous mixing structure within the metropolitan and province groups, or any spatial aspects. The short-term visitors are assumed to have a higher rate of contacts with provincial people (staying at hotels and eating at restaurants) but it is not assumed that they also have higher mixing rates within the metropolitan group. Further, a conservative assumption is that the mixing contacts are \emph{added} to the contacts within the group (rather than \emph{replacing} such contacts which would result in fewer infections).

We have not found much related work addressing similar type of questions. More common is to study effects of more frequent movements between regions, for example movement of people between smaller towns and a larger urban city \cite{AP15}. Such scenario would correspond more to the situation with only having short-term visitors and no long-term visitors.

\section*{Acknowledgements}

We thank Cecilia Thomsson at Gotlands Allehanda (newspaper on Gotland) for supplying details to the Gotland example. T.B.\ is grateful to the Swedish Research Council for financial support.

\newpage

\section*{Supplementary information}

\subsection*{Model details: final size considerations}

We outline the underlying multitype SIR epidemic model and corresponding final size results, using broadly the notation of~\cite{AB00}, Chapter 6,  which should be consulted for further details.  Suppose that the population is split into $k$ groups (types) of individuals labelled $1,2,\dots,k$.  In the model in the main text, $k=3$, type $1$ is metropolitan short-term, type $2$ is metropolitan long-term and type $3$ is provincial.  Let $n_i$ denote the size of group $i$ ($i=1,2,\dots,k$) and $n=n_1+n_2+\dots+n_k$ denote the total population size.  For $i=1,2,\dots, k$, let $\pi_i=n_i/n$ denote the fraction of the population that is of type $i$.  Infective individuals have independent infectious periods, which, for simplicity, follow the same distribution irrespective of type.  The final size results we describe depend on that distribution only through its mean, which we denote by $\iota$.  These final size results are also invariant to the introduction of an exposed (latent) period into the model, so they hold for a corresponding SEIR model.  We assume that a given type-$i$ infective transmits infection to a given type-$j$
susceptible at rate $\lambda_{ij}/n$ (more precisely at the points of a Poisson process having rate $\lambda_{ij}/n$); i.e.~the individual-to-individual infection rate between a type-$i$ infective and a type-$j$ susceptible is $\lambda_{ij}/n$.  The epidemic starts with a small number of initial infectives, with the rest of the population being susceptible, and ends when there is no infective individual in the population.  We assume that the size of \emph{each} group is large, so the total population size $n$ is also large.

For $i,j=1,2,\dots,k$, let $m_{ij}$ be the mean number of type-$j$ susceptibles infected by a typical type-$i$ infective at the start of the epidemic.  Then
\begin{equation}
\label{equ:NGM}
m_{ij}=\iota \frac{\lambda_{ij}}{n} n_j =  \iota \lambda_{ij} \pi_j \qquad (i,j=j=1,2,\dots,k).
\end{equation}
Let $M$ be the $k \times k$ matrix whose $(i,j)$th element is $m_{ij}$.  The matrix $M$ is called the
next-generation matrix.  The basic reproduction number $R_0$ is given by the maximal eigenvalue of $M$.
If $R_0 \le 1$ then the probability that an epidemic with few initial infectives takes off and leads to a major outbreak is $0$ in the limit as $n \to \infty$.  If $R_0>1$ then there is a non-zero probability that a major outbreak occurs.

Suppose that $R_0>1$ and a major outbreak occurs.  For $i=1,2,\dots,k$, let $Z_i$ be the total number of individuals in group $i$ that are infected by the epidemic and $\bar{Z}_i=Z_i/n_i$ be the fraction of group $i$ that is infected.  Then, for large $n$, $\bar{Z}_i \approx \tau_i$, where $(\tau_1, \tau_2, \dots, \tau_k)$ is
the unique non-zero solution of the equations
\begin{equation}
\label{equ:tau}
1-\tau_j={\rm e}^{-\iota \sum_{i=1}^k \pi_i \tau_i \lambda_{ij}}\qquad (j=1,2,\dots, k).
\end{equation}
This approximation becomes exact in the limit $n \to \infty$ in that $(\bar{Z}_1, \bar{Z}_2, \dots, \bar{Z}_k)$
converges in probability to $(\tau_1, \tau_2, \dots, \tau_k)$ as $n \to \infty$.

Suppose now that at the start of the epidemic a fraction $r_i$ of individuals in group $i$ are immune to infection ($i=1,2,\dots,k$).  Thus, the number of susceptibles in group $i$ is $n_i'=(1-r_i)n_i$ $(i=1,2,\dots,k)$ and the total number of susceptibles is $n'=n_1(1-r_1)+n_2(1-r_2)+\dots+n_k(1-r_k)$.
Hence, the fraction of the susceptible population that belong to group $i$ is now
\begin{equation}
\label{equ:pi'}
\pi_i'=\frac{\pi_i(1-r_i)}{\sum_{j=1}^k  \pi_j(1-r_j)}\qquad (i=1,2,\dots,k).
\end{equation}
The individual-to-individual infection rates remain unchanged so, in an obvious notation,
$\lambda_{ij}'/n'=\lambda_{ij}/n$, whence
\begin{equation}
\label{equ:lambda'}
\lambda_{ij}'=\lambda_{ij} \sum_{l=1}^k  \pi_l(1-r_l) \qquad (i,j=j=1,2,\dots,k).
\end{equation}
Substituting~\eqref{equ:pi'} and~\eqref{equ:lambda'} into~\eqref{equ:NGM}, it follows that the basic reproduction number is now $R_0'$, where $R_0'$ is the maximal eigenvalue of
the matrix $M'$ with elements given by
\begin{equation}
\label{equ:NGM'}
m_{ij}'=\iota \lambda_{ij}' \pi_j'=\iota \lambda_{ij} \pi_j (1-r_j) \qquad (i,j=j=1,2,\dots,k).
\end{equation}

Suppose that $R_0'>1$ and a major outbreak occurs.  For $i=1,2,\dots,k$, let $Z_i'$ be the total number of individuals in group $i$ that are infected by the epidemic and $\bar{Z}_i=Z_i'/n_i'$ be the fraction of
the initially susceptible population in group $i$ that are infected.
Then it follow using~\eqref{equ:tau} that, for large $n$, $\bar{Z}_i' \approx \tau_i'$, where $(\tau_1', \tau_2', \dots, \tau_k')$ is
the unique non-zero solution of the equations
\begin{align}
\label{equ:tau1}
1-\tau_j'&={\rm e}^{-\iota \sum_{i=1}^k \pi_i' \tau_i' \lambda_{ij}'},\nonumber\\
&={\rm e}^{-\iota \sum_{i=1}^k \pi_i(1-r_i) \tau_i' \lambda_{ij}} \qquad (i,j=j=1,2,\dots,k),
\end{align}
using~\eqref{equ:pi'} and~\eqref{equ:lambda'}.

Let $\hat{\tau}_i$ be the limiting (as $n \to \infty$) fraction of the population in group $i$, including immune individuals, that are infected by a major outbreak.  Then $\hat{\tau}_i=(1-r_i) \tau_i'$ ($i=1,2,\dots,k$) and
it follows from~\eqref{equ:tau1} that
\[
1-\frac{\hat{\tau}_j}{1-r_j}={\rm e}^{-\iota \sum_{i=1}^k \pi_i \hat{\tau}_i \lambda_{ij}}\qquad (j=1,2,\dots, k).
\]
Thus, although $R_0'$ can be obtained by simply replacing $\lambda_{ij}$ by $\lambda_{ij}(1-r_j)$ in the next-generation matrix $M$, making this substitution in the equations governing $(\tau_1, \tau_2, \dots, \tau_k)$ typically does not yield $(\hat{\tau}_1, \hat{\tau}_2, \dots, \hat{\tau}_k)$.

\subsection*{Model details: time evolving epidemic}
Consider the SEIR (susceptible $\to$ exposed $\to$ infective $\to$ recovered) version of the above model.  Assume that the exposed and infectious periods follow exponential distributions with means $\nu^{-1}$ and $\gamma^{-1}$, respectively.  For $t \ge 0$ and $j=1,2,\dots,k$, let
$S_j(t), E_j(t)$ and $I_j(t)$ denote respectively the numbers of susceptible, exposed and infective individuals in group $j$ at time $t$.  Further, let $s_j(t)=S_j(t)/n, e_j(t)=E_j(t)/n$ and $i_j(t)=I_j(t)/n$.  In a deterministic formulation, the rate of change in $S_j(t)$ at time $t$ owing to infection is $-S_j(t) \sum_{l=1}^k I_l(t)\lambda_{lj}/n$, so the corresponding rate of change in $s_j(t)$ is $-s_j(t) \sum_{l=1}^k i_l(t)\lambda_{lj}$.  Similarly, the rate of change in $e_j(t)$ owing to exposed type-$j$ individuals becoming infective is $-\nu e_j(t)$ and the rate of change in $i_j(t)$ owing to type-$j$ infectives recovering (and becoming immune) is $-\gamma i_j(t)$.  We also allow individuals to both leave and enter the population.  Specifically, a type-$j$ individual leaves at rate $\mu_j$, irrespective of disease status, with corresponding rates of change in $s_j(t), e_j(t)$ and $i_j(t)$ of $-\mu_j s_j(t), -\mu_j e_j(t)$ and $-\mu_j i_j(t)$, respectively.  Type-$j$ individuals enter the population at the same overall rate as they leave.  Each entering type-$j$ individual is independently susceptible, exposed or infective with probabilities $p_{jS}, p_{jE}$ and
$p_{jI}$, respectively, leading to rates of change in $s_j(t), e_j(t)$ and $i_j(t)$ of $\mu_j\pi_j p_{jS}, \mu_j \pi_j p_{jE}$ and $\mu_j \pi_j p_{jI}$, respectively.  Note that $p_{jS}+p_{jE}+p_{jI}$ may be $<1$ as incoming
individuals may also be recovered.  Putting all of this together leads to the following system of ordinary differential equations (where $j=1,2,\dots, k$):
\begin{align*}
\dfrac{ds_j}{dt}&=-s_j \sum_{l=1}^k i_l\lambda_{lj}-\mu_j s_j+\mu \pi_j p_{jS},\\
\dfrac{de_j}{dt}&= s_j \sum_{l=1}^k i_l\lambda_{lj}-\nu e_j-\mu_j e_j+\mu \pi_j p_{jE},\\
\dfrac{di_j}{dt}&=\nu e_j-\gamma i_j-\mu_j i_j+\mu \pi_j p_{jI}.
\end{align*}

For given initial condition $(s_j(0), e_j(0), i_j(0))$ $(j=1,2,\dots,k)$ satisfying $\sum_{j=1}^k (e_j(0)+i_j(0))>0$, the solution of the above system of differential equations approximates closely the
behaviour of the corresponding Markovian stochastic model for large $n$; see~\cite{EK86}, Chapter 11, and~\cite{AB00}, Chapter 5, for further details.  Note that the condition $\sum_{j=1}^k (e_j(0)+i_j(0))>0$ implies that initially a strictly positive fraction of the population is infective or exposed in the limit as $n \to \infty$.  Note also that the initial condition must satisfy $0 \le s_j(0)+e_j(0)+i_j(0)\le \pi_j$ for $j=1,2,\dots,k$.

In the above it is assumed that the infection rates etc.~are not time-dependent.  If, for example, the infection rates and migration rates are time-dependent, then, in an obvious notation, the above set of differential equations is replaced by:

\begin{align*}
\dfrac{ds_j(t)}{dt}&=-s_j(t) \sum_{l=1}^k i_l(t)\lambda_{lj}(t)-\mu_j(t) s_j(t)+\mu(t) \pi_j p_{jS},\\
\dfrac{de_j(t)}{dt}&= s_j(t) \sum_{l=1}^k i_l(t)\lambda_{lj}(t)-\nu e_j(t)-\mu_j(t) e_j(t)+\mu(t) \pi_j p_{jE},\\
\dfrac{di_j(t)}{dt}&=\nu e_j(t)-\gamma i_j(t)-\mu_j(t) i_j(t)+\mu(t) \pi_j p_{jI}.
\end{align*}

\subsection*{Connection to notation in main text}
\subsubsection*{Next-generation matrix}
We assume that the mean of the infectious period $\iota=\gamma^{-1}$.  
Recall that $k=3$, type $1$ is metropolitan short-term, type $2$ is metropolitan long-term and type $3$ is provincial.  Equating the $(3,3)$ element of the
next-generation matrix $M'$ at~\eqref{equ:NGM'} with the corresponding element of the matrix $A$ in the main text and recalling that $R_E^{(P)}=R_0^{(P)}(1-c^{(P)})(1-r^{(P)})$ yields $\gamma^{-1}\lambda_{33}(1-\pi_M)=R_0^{(P)}(1-c^{(P)})$, so $\lambda_{33}=\gamma R_0^{(P)}(1-c^{(P)})/(1-\pi_M)$.  A similar calculation shows that
\[
\lambda_{11}=\lambda_{22}=\lambda_{21}=\lambda_{22}=R_0^{(M)}(1-c^{(M)})/\pi_M.
\]
Recall that $R_{M\to P}$ is the mean number an infected long-term $M$-individual has with $P$-individuals during
his/her infectious period.  Thus, setting $i=2$ and $j=3$ in~\eqref{equ:NGM}, $R_{M\to P}=\gamma^{-1}\lambda_{23}(1-\pi_M)$, so $\lambda_{23}=\gamma R_{M\to P}/(1-\pi_M)$.  Further, $\lambda_{32}=\gamma R_{M\to P}/(1-\pi_M)$, as contacts are symmetric.  Finally, 
\[
\lambda_{13}=\lambda_{31}=\gamma (1+k_S)R_{M\to P}/(1-\pi_M),
\]
since short-term metropolitan visitors have on average $1+k_S$ times as many contacts with provincials
as do long-term visitors. Thus,
\begin{equation}
\label{equ:Lambda}
\Lambda=
\begin{pmatrix}
\gamma R_0^{(M)}(1-c^{(M)})/\pi_M & \gamma R_0^{(M)}(1-c^{(M)})/\pi_M& \gamma (1+k_S) R_{M \to P}/(1-\pi_M)\\
\gamma R_0^{(M)}(1-c^{(M)})/\pi_M & \gamma R_0^{(M)}(1-c^{(M)})/\pi_M& \gamma R_{M \to P}/(1-\pi_M)\\
\gamma (1+k_S) R_{M \to P}/(1-\pi_M)& \gamma R_{M \to P}/(1-\pi_M) & \gamma R_0^{(P)}(1-c^{(P)})/(1-\pi_M)
\end{pmatrix}.
\end{equation}

Recall that a fraction $\alpha_S$ of metropolitan visitors are short-term visitors and
that $\alpha_L=1-\alpha_S$. Thus, $\pi_1= \alpha_S \pi_M, \pi_2= \alpha_L \pi_M$ and
$\pi_3= (1-\pi_M)$.  Suppose that the fraction of metropolitan visitors that are recovered
on arrival is independent of their length of stay.  Then
it follows using~\eqref{equ:NGM'} and~\eqref{equ:Lambda} that the next-generation matrix is
\[
A=
\begin{pmatrix}
 R_E^{(M)}\alpha_S &  R_E^{(M)}\alpha_L&  (1+k_S) R_{M \to P}(1-r^{(P)})\\
 R_E^{(M)}\alpha_S &  R_E^{(M)} \alpha_L&  R_{M \to P}(1-r^{(P)})\\
 (1+k_S) R_{M \to P}(1-r^{(M)})\frac{\alpha_S \pi_M}{1-\pi_M}&  R_{M \to P}(1-r^{(M)})\frac{\alpha_L\pi_M}{1-\pi_M} &  R_E^{(P)}
\end{pmatrix}.
\]

We now derive the upper bound for $R_{M \to P}$, given in the main text,  so that the effective reproduction number $R_E$ is strictly less than $1$.  Recall that $R_E$ is the maximal eigenvalue
of $A$. Thus $f(R_E)=1$, where
$f(\lambda)=|A-\lambda I|$ is the characteristic polynomial of $A$.  Expanding the determinant $|A-\lambda I|$
yields
\begin{align}
\label{equ:flambda}
f(\lambda)=&(R_E^{(M)}\alpha_S -\lambda)\left[(R_E^{(M)} \alpha_L-\lambda)(R_E^{(P)}-\lambda)-
R_{M \to P}^2 (1-r^{(P)})(1-r^{(M)})\frac{\alpha_L\pi_M}{1-\pi_M}\right]\nonumber\\
&-R_E^{(M)}\alpha_L\left[ R_E^{(M)}\alpha_S(R_E^{(P)}-\lambda)-(1+k_S)R_{M \to P}^2
(1-r^{(P)})(1-r^{(M)})\frac{\alpha_S \pi_M}{1-\pi_M}\right]\nonumber\\
&+(1+k_S)R_{M \to P}(1-r^{(P)})\left[R_E^{(M)}\alpha_S R_{M \to P}(1-r^{(M)})\frac{\alpha_L\pi_M}{1-\pi_M}\right.\nonumber\\
&\left.\qquad\qquad\qquad\qquad\qquad\qquad-(R_E^{(M)} \alpha_L-\lambda)(1+k_S) R_{M \to P}(1-r^{(M)})\frac{\alpha_S \pi_M}{1-\pi_M}\right].
\end{align}

Assume that $r^{(M)}<1$ and
$r^{(P)}<1$, since otherwise at least one of the populations is fully immune.  Assume also that
$R_{M \to P}>0$, since otherwise there is no interaction between the two populations.
Under these assumptions, clearly $R_E> 1$ if either $R_E^{(M)}\ge 1$ or $R_E^{(P)} \ge 1$, so assume that
$R_E^{(M)}$ and $R_E^{(P)}$ are both $<1$.
We determine the greatest value of $R_{M \to P}$ so that $R_E=1$, assuming that all other parameters are held fixed. A necessary condition for
$R_E=1$ is that $f(1)=0$ and hence, using~\eqref{equ:flambda} and a little algebra, that
\begin{equation}
\label{equ:RE=1}
R_{M \to P}^2 (1-r^{(P)})(1-r^{(M)})\frac{\pi_M}{1-\pi_M}[(1+k_S \alpha_S)^2+\alpha_S(1-\alpha_S)(1-R_E^{(M)})]
=(1-R_E^{(M)})(1-R_E^{(P)}).
\end{equation}

Now $R_E$ is strictly increasing in $R_{M \to P}$, as all elements of $A'$ are nonnegative and increasing in $R_{M \to P}$, and at least one element is strictly increasing in $R_{M \to P}$.  Further, $R_E<1$ for all sufficiently small $R_{M \to P}$ and $R_E \to \infty$ as $R_{M \to P} \to \infty$.  It follows that
there exists a unique value of $R_{M \to P}$ such that $R_E=1$ and, using~\eqref{equ:RE=1},  that
$R_E \le 1$ if and only if
\[
R_{M \to P} \le \sqrt{\frac{(1-R_E^{(M)})(1-R_E^{(P)})(1-\pi_M)}
{(1-r^{(P)})(1-r^{(M)})\pi_M[(1+k_S \alpha_S)^2+\alpha_S(1-\alpha_S)(1-R_E^{(M)})]}}.
\]

\subsubsection*{Temporal dynamics}

For the time evolving epidemic we assume that in the provincial population initially a fraction $\epsilon^{(P)}$ is infected (either exposed or infective), a fraction $r^{(P)}$ is recovered
(and immune) and the remaining fraction $1-\epsilon^{(P)}-r^{(P)}$ is susceptible.  The corresponding quantities for the metropolitan population are assumed to be $\epsilon^{(M)}, r^{(M)}$ and $1-\epsilon^{(M)}-r^{(M)}$ for both short-term and long-term visitors.
For the initially infected we assume that a fraction $\theta_E=\nu^{-1}/(\nu^{-1}+\gamma^{-1})=\gamma/(\nu+\gamma)$ is exposed, so a fraction $\theta_I=\nu/(\nu+\gamma)$
is infective.  These assumptions yield the following initial condition for the system of differential equations:
\begin{align*}
(s_1(0), e_1(0), i_1(0))&=\alpha_S \pi_M (1-\epsilon^{(M)}-r^{(M)}, \theta_E \epsilon^{(M)}, \theta_I \epsilon^{(M)}),\\
(s_2(0), e_2(0), i_2(0))&=\alpha_L \pi_M (1-\epsilon^{(M)}-r^{(M)}, \theta_E \epsilon^{(M)}, \theta_I \epsilon^{(M)}),\\
(s_3(0), e_3(0), i_3(0))&=(1-\pi_M) (1-\epsilon^{(P)}-r^{(P)}, \theta_E \epsilon^{(P)}, \theta_I \epsilon^{(P)}),\\
\end{align*}

Finally, since only short-term metropolitan visitors leave the population during the course of the study, $\mu_1=\mu_S, \mu_2=0$ and $\mu_3=0$.  We assume that the fractions of arriving short-term metropolitan visitors that are susceptible, exposed and infective are the same as in that population at time $t=0$, so $p_{1S}= 1-\epsilon^{(M)}-r^{(M)}, p_{1E}= \theta_E \epsilon^{(M)}$ and $p_{1I}= \theta_I \epsilon^{(M)}$.

The situation studied in the main paper, in which all metropolitan individuals leave at time $t_0=35$ days,
can be analysed by setting, for all $i,j=1,2,3$,
\[
\lambda_{ij}(t)=
\begin{cases}
	      \lambda_{ij}& \text{ if } t <t_0, \\
	      \lambda_{ij}' & \text{ if } t \ge t_0,
\end{cases}
\]
where $\lambda_{ij}$ is given by~\eqref{equ:Lambda}, $\lambda_{33}'=\lambda_{33}$ and
$\lambda_{ij}'=0$ for all other $(i,j)$.

\begin{figure}[ht]
\begin{center}
\includegraphics[scale=0.9, angle=0]{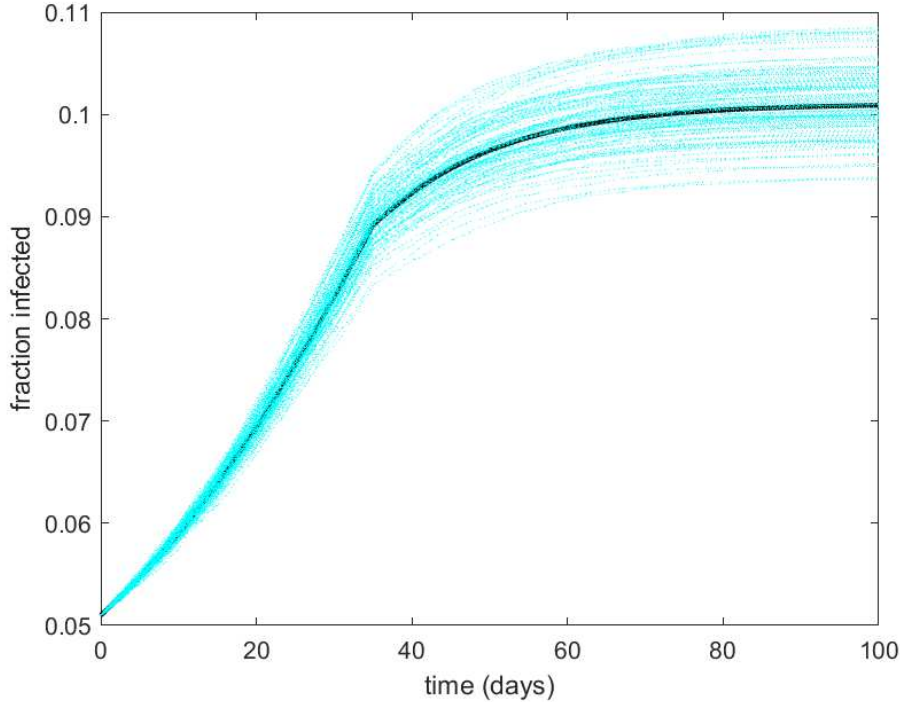}
\end{center}
\caption{Plot of cumulative fraction infected among provincial people over time (starting with 5\% immunity) for the deterministic model (solid line), with $100$ simulated realisations of the corresponding Markovian stochastic model
superimposed (dotted lines).  Metropolitans visit the  province for 35 days, after which only provincial people remain.  }
\label{Fig_timeevolve1}
\end{figure}

Figure~\ref{Fig_timeevolve1} shows the time evolution of the cumulative fraction of $P$-individuals infected
for the deteministic model, with $100$ simulations of the corresponding Markovian stochastic model
superimposed.  The parameters are the same as those for the base model in Figure~\ref{Fig_timeevolve} in the main text.  For the stochastic simulations, the population consists of $60 000$ provincial individuals, $50 000$ short-term metropolitan visitors and $100 000$ long-term metropolitan visitors.  The deterministic trajectory clearly approximates very well the mean of the stochastic trajectory.  The variability of the stochastic trajectories about the deterministic trajectory would be reduced if the population sizes were increased.

\subsection*{Relative effects of $R_0^{(M)}$ and $R_0^{(P)}$ on the final fraction infected}
We investigate the apparent discrepancy between the relative effects of $R_0^{(M)}$ and $R_0^{(P)}$ in
Figure~\ref{Fig_R0} and Table~\ref{tab_frac_inf} in the main text.  Recall that $R_0^{(M)}$ has a greater impact than $R_0^{(P)}$ on the effective reproduction number $R_E$ in Figure~\ref{Fig_R0} while, in Table~\ref{tab_frac_inf}, the effect of $R_0^{(P)}$ on the final fraction infected in the provincial population
is appreciably larger than that of $R_0^{(M)}$.

\begin{figure}[ht]
\begin{center}
\includegraphics[scale=0.9, angle=0]{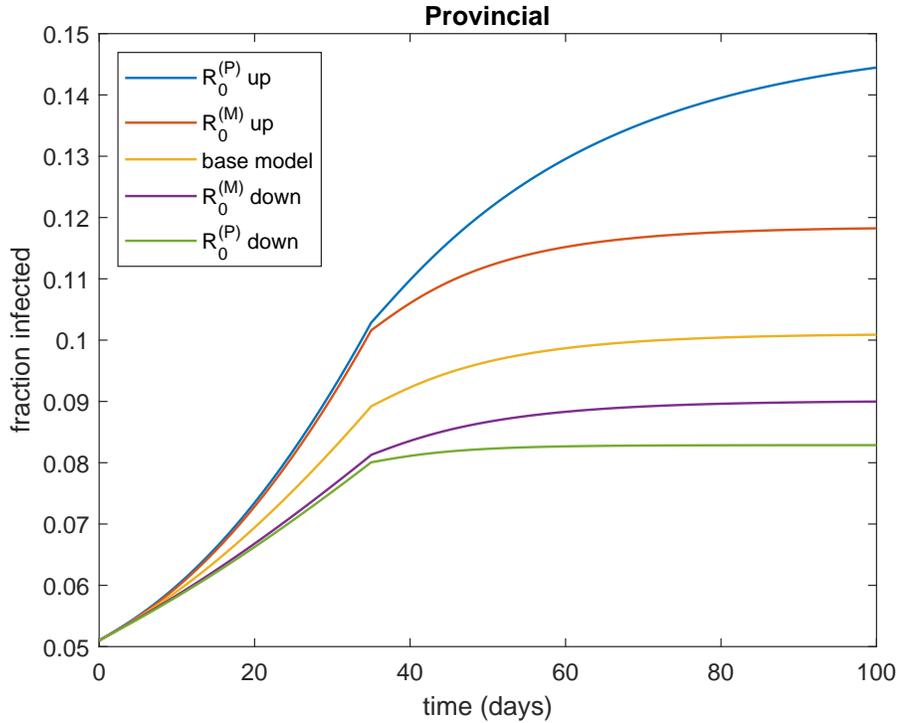}
\end{center}
\caption{Plot of cumulative fraction infected among provincial people over time (starting with 5\% immunity) for the base model and for the models in which the parameters $R_0^{(M)}$ and $R_0^{(P)}$ are varied individually;  `up' means that parameter is increased by $50\%$, with all other parameters held fixed, and `down' means that it is decreased by $50\%$. }
\label{Fig_R0MR0P1}
\end{figure}

In Figure~\ref{Fig_R0MR0P1} we plot the time evolution of the epidemic within the provincial population
under the different values of $R_0^{(M)}$ and $R_0^{(P)}$ considered in Table~\ref{tab_frac_inf}.  It is evident that there is little difference between the impacts of changing $R_0^{(M)}$ and changing $R_0^{(P)}$ over the period when the metropolitan people are on the island and that the large impact of $R_0^{(P)}$ on the final fraction infected on the island is owing to the further spread on the island after the metropolitan visitors have left.  Of course, the value of $R_0^{(M)}$ has no effect on the spread after day 35.

\begin{figure}[ht]
\begin{center}
\includegraphics[scale=0.9, angle=0]{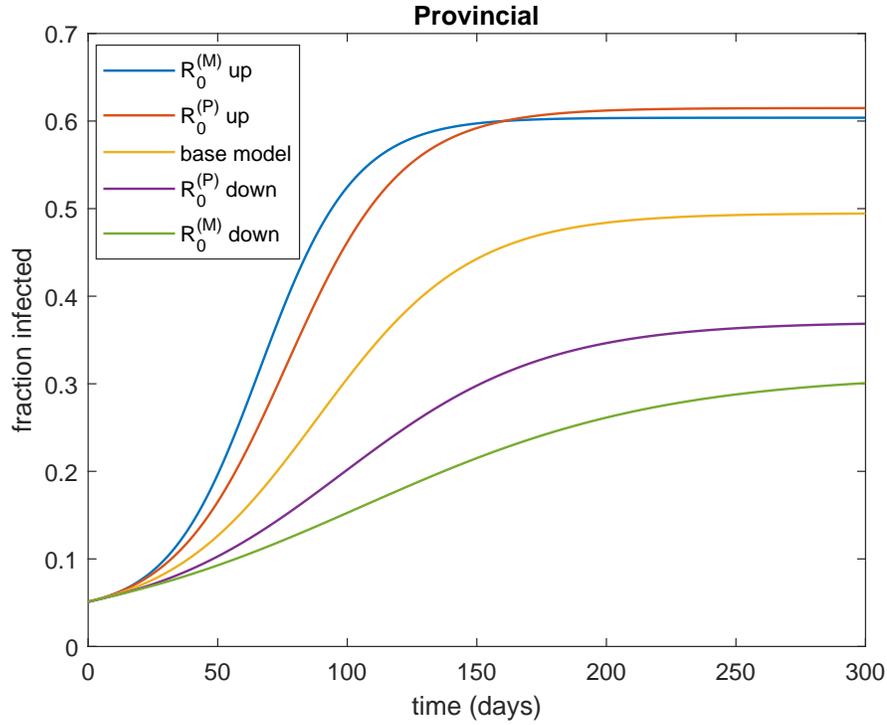}
\end{center}
\caption{Plot of cumulative fraction infected over time (starting with 5\% immunity) for the base model and for the models in which the parameters $R_0^{(M)}$ and $R_0^{(P)}$ are varied individually, assuming that metropolitan visitors stay forever;  `up' means that parameter is increased by $50\%$, with all other parameters held fixed, and `down' means that it is decreased by $50\%$. }
\label{Fig_R0MR0P2}
\end{figure}

\begin{figure}
\begin{center}
\resizebox{\hfigwidth}{!}{\includegraphics{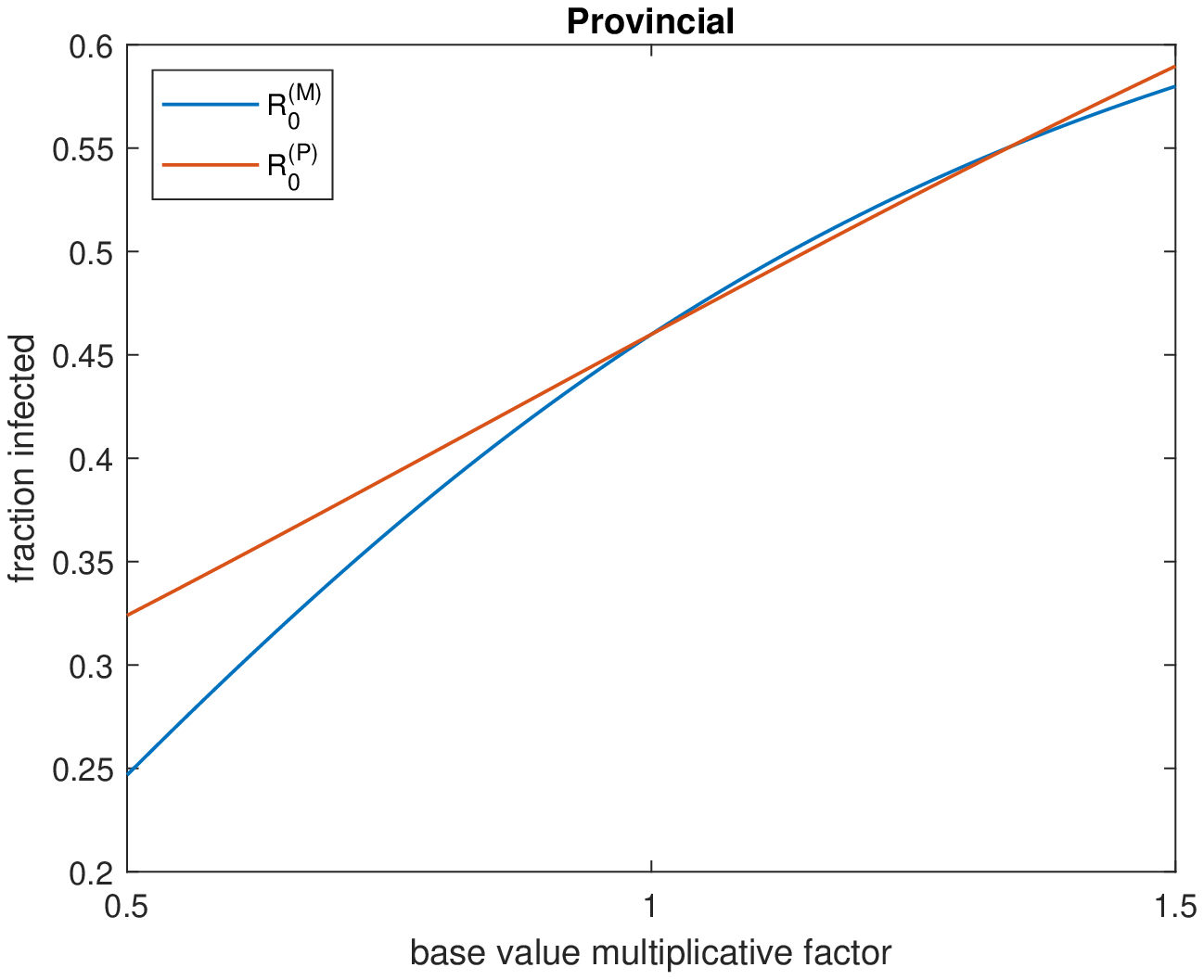}}
\resizebox{\hfigwidth}{!}{\includegraphics{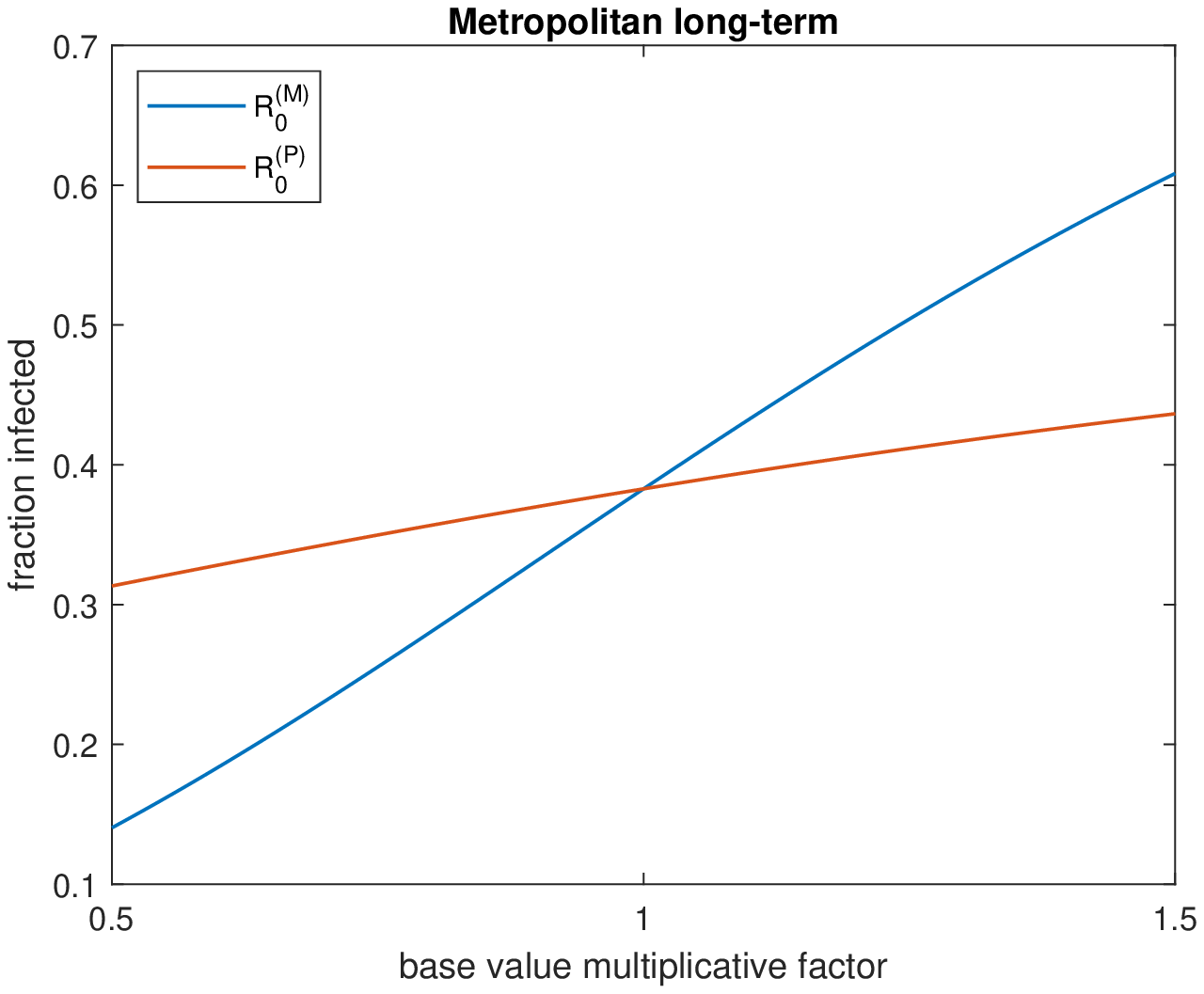}}\\
\resizebox{\hfigwidth}{!}{\includegraphics{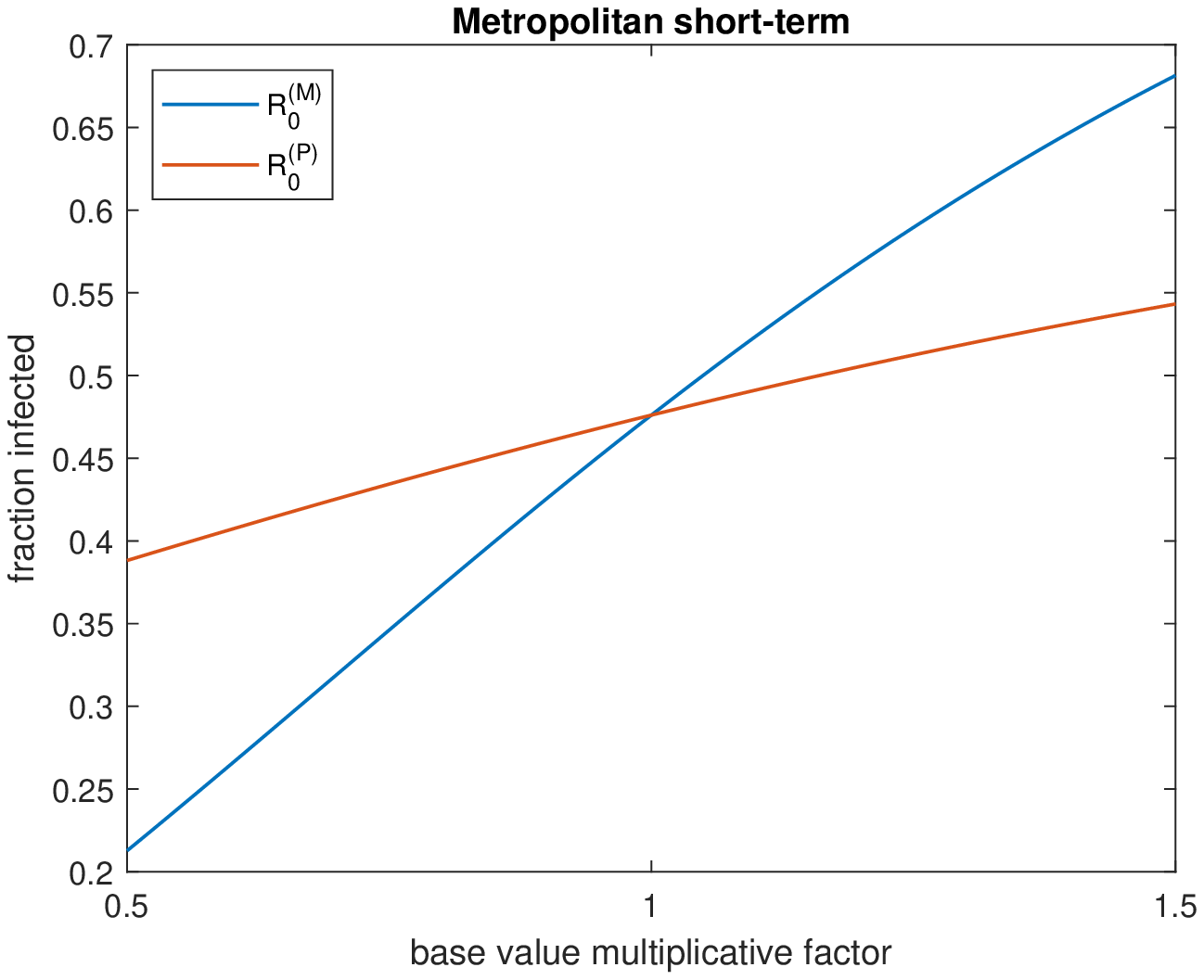}}
\resizebox{\hfigwidth}{!}{\includegraphics{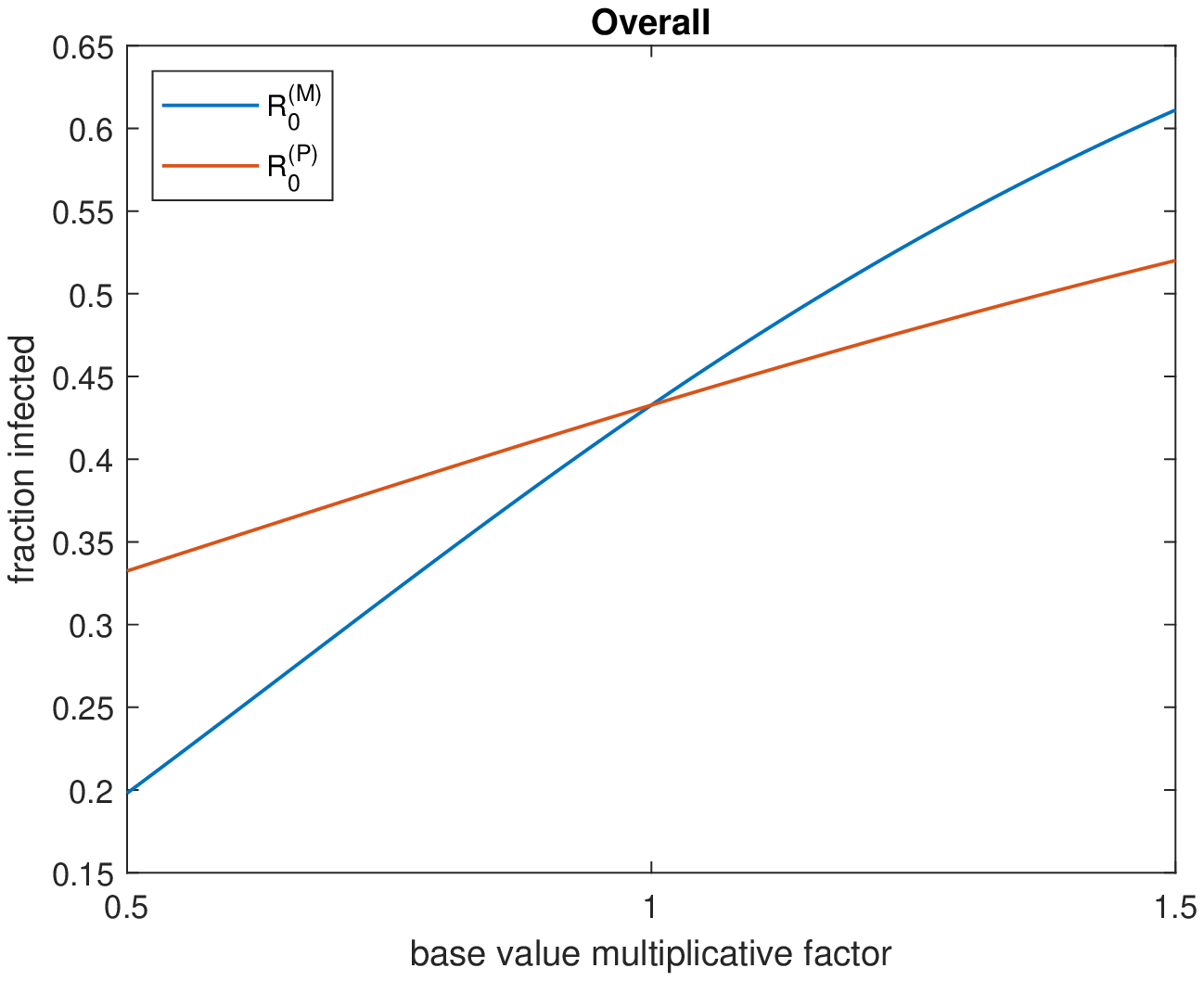}}
\end{center}
\caption{Plots of final fraction infected  where each of $R_0^{(M)}$ and $R_0^{(P)}$ are varied individually.  The final fractions infected in the provincial, long-term metropolitan and short-term metropolitan populations are shown in the top-left, top-right and bottom-left panels, respectively, while the overall final fraction infected is shown in the bottom-right panel.}
\label{Fig_R0MR0P3}
\end{figure}

In Figure~\ref{Fig_R0MR0P2}we give similar plots but assuming that the metropolitan people stay on the island indefinitely, so there is no migration.  Note that now $R_0^{(M)}$ has a greater influence than $R_0^{(P)}$
throughout most of the epidemic; indeed it is only near the end of the epidemic that increasing $R_0^{(P)}$ by
$50\%$ has a greater effect than increasing $R_0^{(M)}$ by the same amount.  This is explored further in Figure~\ref{Fig_R0MR0P3}, which shows the effects of $R_0^{(M)}$ and $R_0^{(P)}$ on the final fraction infected in each of the sub-populations and overall.   Note that while large reductions and increases in $R_0^{(P)}$ are more influential than corresponding changes in $R_0^{(M)}$ for the
provincial population, that is not the case for the two metropolitan populations, for which $R_0^{(M)}$ has an across-the-board greater effect than $R_0^{(P)}$.  This is also the case for the population as a whole, which is fully consistent with the relative effects of $R_0^{(M)}$ and $R_0^{(P)}$
on the effective reproduction $R_E$ depicted in Figure~\ref{Fig_R0} in the main paper.  

\end{document}